\documentclass[letterpaper]{svjour2}
\smartqed
\newcommand{\ve}[1]{{\rm{\bf {#1}}}}

\newcommand{\dep}{(a,u)}
\newcommand{\depp}{(a+da,u+du)}

\usepackage{graphicx}
\usepackage{amssymb}
\usepackage{amsmath}
\usepackage{rotate}
\begin{document}

\title{VFISV: Very Fast Inversion of the Stokes Vector for the
Helioseismic and Magnetic Imager.}

\author{J.M.~Borrero \and S.~Tomczyk \and M.~Kubo \and H.~Socas-Navarro \and J.~Schou \and
S.~Couvidat \and R.~Bogart}

\institute{J.M.~ Borrero, S.~Tomczyk, M.~Kubo, H. Socas-Navarro \at
                 High Altitude Observatory, National Corporation for \\
                 Atmospheric Research, 3080 Center Green CG-1, Boulder CO 80301, USA\\
                 \email{borrero@ucar.edu, tomczyk@ucar.edu, kubo@ucar.edu, navarro@ucar.edu}\\
		 \and
		 H.~Socas-Navarro \at Instituto de Astrof{\'\i}sica de Canarias, \\
		 Avda V{\'\i}a L\'actea S/N, La Laguna 38205, Tenerife, Spain\\
		 \email{hsocas@iac.es}\\
		 \and
           J.~Schou S.~Couvidat, R.~Bogart \at
	         Hansen Experimental Physics Laboratory, Stanford University \\
		 Stanford, CA 94305, USA\\
		 \email{schou@stanford.edu, rick@stanford.edu, couvidat@stanford.edu}}

\date{Received: date / Accepted: date}
\maketitle
\begin{abstract}
In this paper we describe in detail the implementation and main properties
of a new inversion code for the polarized radiative transfer equation (VFISV: Very Fast 
inversion of the Stokes vector). VFISV will routinely analyze pipeline
data from the Helioseismic and Magnetic Imager (HMI) on-board of the Solar 
Dynamics Observatory (SDO). It will provide full-disk maps (4096$\times$4096 pixels) 
of the magnetic field vector on the Solar Photosphere every 10 minutes. For this reason 
VFISV is optimized to achieve an inversion speed that will allow it to invert 16 
million pixels every 10 minutes with a modest number (approx. 50) of CPUs.
Here we focus on describing a number of important details, simplifications
and tweaks that have allowed us to significantly speed up the inversion process. 
We also give details on tests performed with data from the spectropolarimeter on-board of 
the Hinode spacecraft.
\end{abstract}
%
\section{Introduction and Motivation}%

The Solar Dynamic Observatory will be launched from Cape
Canaveral in June 2009 / 2010 on an Atlas V Booster.
On board this satellite there will be several instruments dedicated
to the study of the Solar photospheric and coronal magnetic fields 
and their relation to the interplanetary medium, space weather and 
Earth climatology. The Helioseismic and Magnetic Imager is
an instrument developed at Stanford University, Lockheed-Martin
Solar and Astrophysics Laboratory and at the High Altitude Observatory.
HMI consists of a combination of a Lyot filter and two Michelson 
interferometers. Between the two Michelson there is a set of three
10th order retarders: $\lambda/2$, $\lambda/4$ and $\lambda/2$. Located after
these, there is a beam splitter that divides the light between two twin 4096$\times$4096
CCD cameras. Each camera will acquire full-disk images of the Sun with a pixel size
of 0.5 arcsec.

The first camera, to be referred to as {\it Doppler camera}, will be devoted to 
the measurement of the right and left circular polarization,
$I \pm V$, at 6 wavelength positions across the Fe I ($g_{\rm eff}$=2.5) 6173.35 \AA~ line
 every 50 seconds or less. With this, full disk maps of the line of sight components 
of the magnetic field  and velocity will be produced with a 50 second cadence for 
helioseismic studies. The second one, hereafter referred to as {\it vector camera}, 
will measure the full Stokes vector $\ve{I} = (I, Q, U, V)$ at the same wavelengths
as the Doppler camera. The cadence in the vector camera will be about 120 seconds.
Borrero et al. (2007) have conducted a detailed study of the vector camera performance, 
and concluded that photon noise and p-mode cross-talk between the different Stokes parameters
makes the 2-minute data inappropriate to determine the magnetic field vector
with the desired degree of accuracy. They suggest averaging the observed
Stokes vector every 10 minutes before it is inverted.

Therefore our objective has been to create an inversion code for the radiative transfer equation
able to process 13.5 million pixels\footnote{The CCD has 16 million pixels, however only about 13.5 million
will be left after removing off-limb pixels.} in 10 minutes: {\it 22500 pixels/second}. Our task benefits
from the fact that each pixel is treated independently, and therefore our problem can be parallelized
in a straight-forward fashion, resulting in a time saving which is directly proportional to the number
of CPUs employed. Consequently, very fast ({\it 3000-10000 pixels/second}) inversion strategies such a 
Principal Component Analysis (PCA; Rees et al. 2000; Socas-Navarro et al. 2001) or Artificial Neural Networks 
(ANNs; Socas-Navarro 2003, 2005) could meet the HMI requirements with a very  small number of CPUs. Without entering into details about
the different methods, their accuracy and robustness (an interested reader is referred to the recent
reviews by Del Toro iniesta 2003a and Bellot Rubio 2006), it is generally acknowledged that these fast inversion techniques
retrieve a magnetic field vector that is more suitable for a qualitative analysis such as active region evolution and 
tracking, global magnetic field appearance. Our aim with HMI is to provide the magnetic field vector on the 
solar surface with an accuracy good enough to carry out more quantitative studies.  A further concern, from the point of
view of HMI characteristics and scientific objectives, is the fact that current implementations of ANNs do not allow
to obtain a measure of the goodness of the inversion, e.g. $\chi^2$. PCA does retrieve such a measure
in terms of the so-called PCA distance. However, Principal Component Analysis is usually employed to reduce the
dimensionality of the problem (e.g. by inverting the first PCA or Fourier coefficients of the observed Stokes vector).
It is unclear how PCA will perform when inverting data from filtergram instruments, such as HMI, where the number of
data points is already very small (six points in wavelength across the four Stokes parameters).

A good compromise between accuracy and speed is achieved by traditional iterative non-linear squares fitting algorithms, 
specially if applied to the Milne-Eddington (M-E) solution of the radiative transfer equation (Landolfi
\& Landi Degl'Innocenti 1982). The reason for this lies in the simplified thermodynamics of this approximation, that avoids tedious iterative
calculations and evaluations of the partial pressures, ionization and hydrostatic equilibrium, etc (Ruiz Cobo 2006).
In addition, the analytical nature of the M-E solution allows to calculate also analytical derivatives, which significantly speeds
up the inversion process. The first M-E inversion code that fully takes into account magneto-optical effects dates back, to the best
of our knowledge, to Auer et al. (1977) (see also Skumanich \& Lites 1987). This has been for many years the standard inversion technique employed in the
analysis of Advance Stokes Polarimeter data (ASP; previously Stokes II: Baur et al. 1981), and thus it is widely known
 as the HAO/ASP inversion code. As a consequence of the increase in the CPUs speed since the late 80's, 
the HAO/ASP code has gone from inverting several pixels per hour to about {\it 10 pixels/second}. Thus, meeting HMI's goal
would require more than 2000 CPUs. Clearly further modifications and refinements over these type of M-E inversion codes
are necessary if we are to meet HMI's requirements with a limited number ($<$60) of CPUs.

After a brief introduction to Stokes inversion (Section 2), we have divided these modifications into three groups:
those made on the synthesis module of the code (Section 3); modifications made on the inversion module of the code 
(Section 4), and implementation of an accurate initial guess model for the inversion process (Section 5). In Section 6
we present a study of the inversion code's profile and speed. Section 7 shows results obtained from the application of 
VFISV to Hinode/SP data and finally, Section 8 summarizes our results.

\section{Introduction to Stokes inversions}%

In this section we will describe very briefly how a Stokes inversion code typically works. The idea is to familiarize the
reader with the common procedures and nomenclature to facilitate the understanding of the following sections. More details
about this topic can be found in (with different levels of complexity) Ruiz Cobo \& Del Toro Iniesta (1992), Frutiger (2001),
Del Toro Iniesta (2003b), Borrero (2004) and references therein.

The basic idea of any Stokes inversion code is to iteratively fit the observed Stokes vector at each
wavelength position $\vec{I}^{\rm obs}(\lambda)=(I,Q,U,V)$\footnote{$I$ is the total intensity, $Q$ and $U$ are the linear
polarization profiles and $V$ is the circular polarization.}. This fit is done by producing a synthetic Stokes
vector $\vec{I}^{\rm syn}(\lambda,\mathcal{M})$, that it is then compared at each wavelength position with the observed one
via the $\chi^2$ of the fit:

\begin{equation}
\chi^2 = \frac{1}{4L-F} \sum_{i=1}^{L}\sum_{j=1}^{4} [I_{j}^{\rm obs}(\lambda_i)-I_{j}^{\rm syn}(\lambda_i,\mathcal{M})]^2 \frac{w_{ij}^2}{\sigma_i^2}
\end{equation}

\noindent where $L$ is the number of wavelength points observed ($L=6$ in HMI's case) and $F$ is the number of free parameters. Therefore
$4L-F$ refers to number of degrees of freedom in the inversion. $I_{j}^{\rm obs}(\lambda_i)$ and $I_{j}^{\rm syn}(\lambda_i)$ refer to each 
of the 4 Stokes parameters ($j$ index) at each wavelength position ($i$ index) for both the observed and synthetic Stokes vector. Section 4.3
discusses in detail the parameters $w_{ij}$ and $\sigma_i$.

The synthetic profiles depend on a series of model parameters $\mathcal{M}=\{M_f\}$ with $f=1,...,F$. These parameters define the physical 
model that we presume as valid: LTE, non-LTE, multi-component atmospheres, gradients along the line-of-sight, etc. In the case of
a Milne-Eddington atmosphere, these typically are: $\mathcal{M}=\{S_0, S_1,\eta_0,a,\Delta\lambda_D,B,\gamma,\Psi,V_{\rm los}, V_{\rm mac},\alpha_{\rm mag}\}$.
The first 5 of them are the thermodynamic parameters: source function at the base of the Photosphere ($S_0$), gradient of the source function ($S_1$), center to
continuum absorption coefficient ($\eta_0$), damping ($a$), and Doppler width of the spectral line ($\Delta\lambda_D$). The next three refer to the three components of the
magnetic field vector: strength $B$, inclination with respect to the observer $\gamma$, and azimuth of the magnetic field vector
in the plane perpendicular to the observer $\Psi$. The following two are kinematic parameters: line-of-sight velocity of the plasma
harboring the magnetic field ($V_{\rm los}$), and the macroturbulent velocity ($V_{\rm mac}$; used to model unresolved velocity fields). 
The last parameter, $\alpha_{\rm mag}$, is a geometrical parameter that defines what portion of the resolution element is filled with a magnetized plasma.

$\vec{I}^{\rm syn}(\lambda,\mathcal{M})$ is computed in the synthesis module of the inversion code. This module also retrieves the $F$ derivatives
of the synthetic Stokes vector with respect to the model parameters (free parameters): $\partial{\vec{I}^{\rm syn}(\lambda,\mathcal{M})}/ \partial{M_f}$.
Once the synthetic Stokes profiles are obtained, we can determine the $\chi^2$ of the fit (Equation 1). With the derivatives it is possible to compute $\nabla\chi^2$,
which is F-dimensional vector where each of its components can be expressed as:

\begin{equation}
\nabla\chi^2_f = \frac{\partial \chi^2(\mathcal{M})}{\partial M_f}
\end{equation}

We can also compute the modified Hessian matrix: $\mathcal{H}_{\rm mod}$. This matrix is basically the regular Hessian matrix, where second
derivatives have been neglected and where the diagonal elements are modified by means of a parameter $\epsilon$:

\begin{eqnarray}
\mathcal{H}_{ij}^{\rm mod} & = \frac{\partial\chi^2}{\partial M_i}\frac{\partial\chi^2}{\partial M_j}  \;\;\; {\rm if}\;\; i \ne j \\
\mathcal{H}_{ii}^{\rm mod} & = (1+\epsilon) \left[\frac{\partial\chi^2}{\partial M_i}\right]^2
\end{eqnarray}

According to the recipe of the Levenberg-Mardquart algorithm, using $\nabla\chi^2$ and $\mathcal{H}_{\rm mod}$ allows us 
to determine the perturbations, $\delta\mathcal{M}$, that must be applied to the model parameters at iteration $h$, $\mathcal{M}_h$, 
in order to improve $\chi^2$.

\begin{equation}
\delta\mathcal{M} = -\mathcal{H}_{\rm mod}^{-1} \nabla\chi^2
\end{equation}

\begin{equation}
\mathcal{M}_{h+1}=\mathcal{M}_h+\delta\mathcal{M}
\end{equation}

To avoid singularities in the inversion of the modified Hessian matrix (Equation 5) the Singular Value Decomposition method
is often employed. Once the new model parameters $\mathcal{M}_{h+1}$ have been determined, the synthesis module is called
again to produce the synthetic Stokes vector and its derivatives, as depicted in Figure 1. Then, a new $\chi^2$ is computed
and the iteration continues.

\begin{figure}
\begin{center}
\includegraphics[width=20cm,angle=-90]{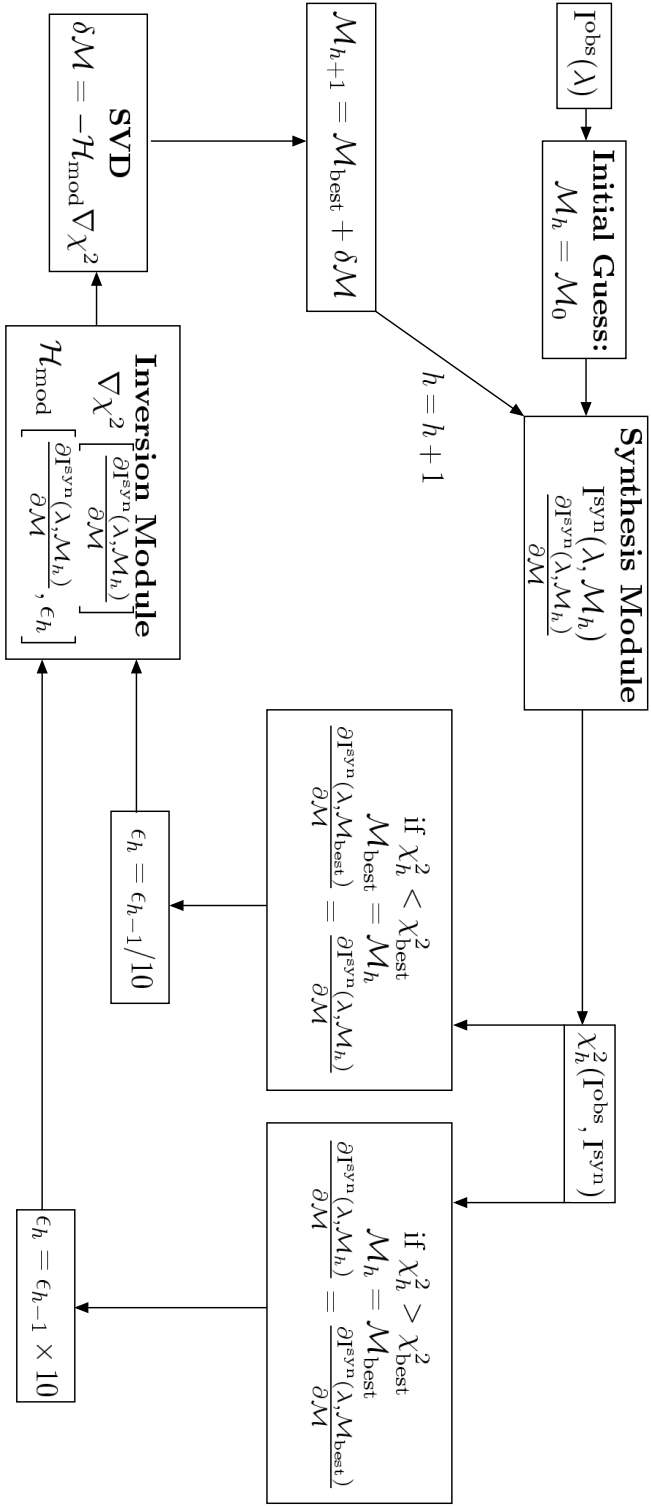}
\caption{Typical iterative scheme for the inversion of Stokes profiles using the Levenberg-Marquadrt algorithm.}
\end{center}
\end{figure}

\section{Synthesis module}%

%
\subsection{Zeeman pattern}

VFISV will be employed for the automatic inversion of data of the Fe I ($g_{\rm geff}$=2.5) 6173.33~\AA~ line (see Table 1).
Therefore a straightforward implementation of its Zeeman pattern can be hardcoded into the inversion code, avoiding 
needless extra calculations using the quantum numbers $L$, $S$ and $J$ of the atomic levels involved.

\begin{table}
\begin{center}
\caption{Atomic parameters for the lower and upper levels of the atomic transition originating the Fe I 6173.33 \AA~
spectral line (from Nave et al. 1994). The Land\'e factors have been calculated in the LS approximation (see for
example Eq.~8.30 in Del Toro Iniesta (2003b).}
\begin{tabular}{|c|c|}
\hline
$\lambda_0$ = 6173.3356 \AA & $g_{\rm eff}$=2.5 \\
\hline
\end{tabular}
\begin{tabular}{|c|c|c|c|c|c|c|}
\hline
Level & Elec. Conf. & $L$ & $S$ & $J$ & Land\'e factor & Excitation Potential [eV]\\
\hline
Lower & $a^5 P$ & 1 & 2 & 1 & $g_l$=2.5 & 2.223 \\
\hline
Upper & $y^5 D^0$ & 2 & 2 & 0 & $g_u$=0 & 4.230 \\
\hline
\end{tabular}
\end{center}
\end{table}

\vspace{0.5cm}
Fe I 6173.33 \AA~ is a Zeeman triplet ($J_l=1 \rightarrow J_u=0$) and therefore its Zeeman pattern
will be composed by one $\pi$ and two $\sigma$ (red and blue) components. In the absence of LOS velocities
the former is unshifted with respect to its central laboratory wavelength $\lambda_0$, whereas the latter are 
shifted (with respect to $\lambda_0$) by an amount given by $\Delta \lambda_B$:

\begin{equation}
\Delta\lambda_B = \pm 4.66685\times 10^{-10} g_{\rm eff} B \lambda_0^2 = \pm 0.044635\times B \;\;\;\; \mbox{[m\AA]}
\end{equation}
\\
\noindent where the positive and negative signs correspond to the red and blue components. $\Delta \lambda_B$ is the only value
that VFISV uses internally. Note that the fact that we are using a Zeeman triplet does not allow VFISV to treat most of the 
magnetic sensitive spectral lines in the solar spectrum. However it is important to point out that some of the most widely 
used ones are indeed also Zeeman triplets: Fe I ($g_{\rm eff}$=3) 5250.2089 \AA~ with $\Delta\lambda_B=0.03859\times B$,
Fe I ($g_{\rm eff}$=2.5) 6302.4936 \AA~ with $\Delta\lambda_B=0.04634\times B$ and Fe I ($g_{\rm eff}$=3) 15648.515 \AA~ with
$\Delta\lambda_B=0.34284\times B$. Therefore, VFISV can be modified to invert those spectral lines by simply changing the
value of $\Delta\lambda_B$.

Having the Zeeman pattern hardcoded inside the code does not increase its speed, 
since this one needs to be calculated only once. The real benefit comes from a simplicity point of view, making the
synthesis, and even more the inversion module, much faster and easier to write and debug.

\subsection{Voigt and Voigt-Faraday functions}

The evaluations of the Voigt and Faraday functions are regarded as one of the most computationally expensive
operations in the synthesis of solar and stellar spectra. A large number of algorithms with different degrees of
sophistication have been proposed, yielding different speeds and accuracies (e.g. Letchworth et al. 2007).
A widely used method is one that obtains the Voigt and Voigt- Faraday functions as the real and imaginary parts of
the quotient of two complex polynomials of a certain order (Hui et al. 1977). This procedure yields very high 
accuracies at reasonable speeds. However, we have found a method, based on the Taylor expansion of the
Voigt and Faraday functions that, although less accurate (but still within our requirements) is significantly
faster. This method takes advantage of the fact that the derivatives of the Voigt $V(a,u)$ and Faraday
$F(a,u)$ functions with respect to the damping $a$ and frequency $u$ 
have the following analytical form (see e.g. Del Toro Iniesta 2003b):

\begin{eqnarray}
\frac{\partial H\dep}{\partial a} =-\frac{\partial F\dep}{\partial u} = -\frac{2}{\sqrt{\pi}}+2aH\dep+2uF\dep = \alpha \\
\frac{\partial H\dep}{\partial u} =\frac{\partial F\dep}{\partial a} = 2aF\dep-2uH\dep = \beta
\end{eqnarray}

With this, it is straightforward to write the second derivatives as:

\begin{eqnarray}
\frac{\partial^2 H\dep}{\partial a^2}=-\frac{\partial^2 H\dep}{\partial u^2}=-\frac{\partial^2 F\dep}{\partial a \partial u} = 2 H\dep+2a\alpha+2u\beta \\
\frac{\partial^2 F\dep}{\partial a^2}=\frac{\partial^2 H\dep}{\partial a \partial u}=-\frac{\partial^2 F\dep}{\partial u^2} = 2 F\dep+2a\beta-2u\alpha
\end{eqnarray}

Let us now create a two-dimensional table with tabulated values for $H$ and $F$ for different pairs of $\dep$. This table can be evaluated 
using any existing algorithm. In our case  we have employed Hui et al. (1977). By knowing $H$ and $F$ at those 
grid points, it is possible to evaluate $H$ and $F$ in the vicinity by extrapolating using a Taylor expansion 
(second order in this example) around the tabulated value:

\begin{align}
H\depp \simeq & H\dep + \alpha da + \beta du + [H\dep+a\alpha+u\beta](da^2-du^2)\notag \\
& +2[F\dep+a\beta-u\alpha]dadu\\
F\depp \simeq & F\dep + \beta da - \alpha du + [F\dep+a\beta-u\alpha](da^2-du^2)\notag \\
& -2[H\dep+a\alpha+u\beta]dadu
\end{align}

For a fixed size of the tabulated values of $H\dep$ and $F\dep$ increasing the order of the Taylor expansion increases of course
the accuracy to which $H\depp$ and $F\depp$ can be determined. The additional number of mathematical operations also increases the computation
time. Figure 2 shows the difference in the determination of $H\depp$ (at $10^6$ frequency $u$ points for a fixed damping of $a=1$) between
our method and the algorithm by  Hui et al. (1977). The upper panel shows the results using only a first order Taylor expansion, whereas the
lower panel was obtained using a second order expansion. The table of tabulated $H\depp$ values had 500 points in frequency and 100 points
in damping. The differences between Hui's algorithm and our approach are always smaller than $10^{-4}$ even for a first order expansion.
As expected the difference drops to zero at the frequencies where the table has been evaluated, and is maximum in between them.
Figure 3 shows the time employed to evaluate the Voigt, $H\dep$ and Voigt-Faraday $F\dep$, functions for different numbers
of frequency points. The comparison reveals that using a second order Taylor expansion decreases the time needed by a factor
or 3.4 as compared to Hui et al. (1977). The improvement reaches a factor 5.4 when using a first order Taylor expansion.

\begin{figure}
\begin{center}
\includegraphics[width=9cm]{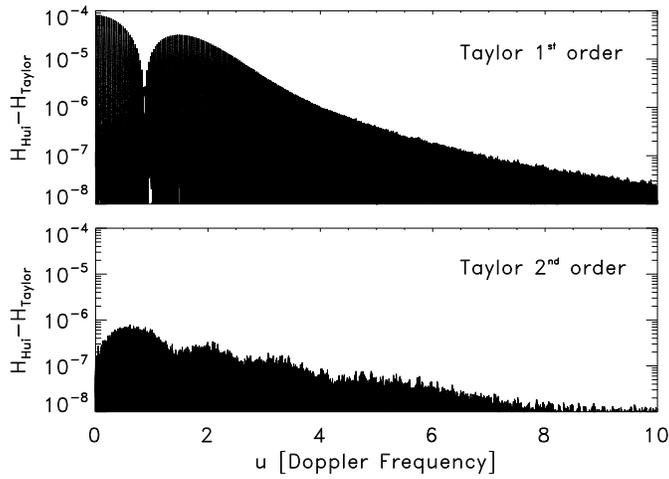}
\caption{{\it Top panel}: differences between the Voigt function calculated using Hui's et al. (1977) algorithm and the Voigt function
calculated with a first order Taylor expansion. In both cases the Voigt function $H\dep$ was evaluated for $10^6$ frequency $u$ points for a
fixed damping $a=1$. {\it Lower panel}: same but this time using a second order Taylor expansion.}
\end{center}
\end{figure}

We have also studied the performance of interpolating $H\depp$ and $F\depp$ in the original table (100 points in damping
and 500 in frequency) using the four nearest tabulated points. It turns out that bilinear interpolation is less accurate 
that using Taylor expansions, because it does not take advantage of Eqs.8 and 9. In addition, it is far slower since it requires
a larger number of evaluations. In fact it is even slower than Hui's et al. algorithm.

\begin{figure}
\begin{center}
\includegraphics[width=9cm]{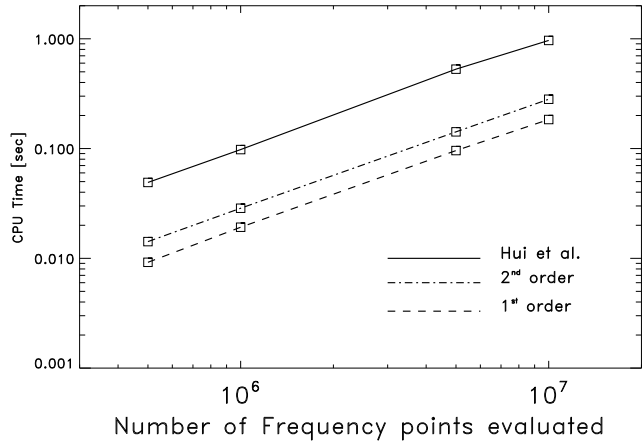}
\caption{Time required to evaluate the Voigt and Faraday functions with increasing number of frequency points. Solid
corresponds to the algorithm by Hui et al. (1977), whereas dashed and dashed-dotted lines correspond to Taylor expansions
of first and second order respectively.}
\end{center}
\end{figure}

\section{Inversion module}%

\subsection{Damping and macroturbulence}

The damping parameter $a$ (see Sect.~2) controls the distance from line center at which the spectral profile
shifts from a Gaussian-like (thermal broadening) to a Lorentzian-like profile (radiative and collisional broadening).
Even in its classical approximation, $a$ is a complicated function of the temperature and partial pressures (\"Unsold 1955,
Wittmann 1974). In the Milne-Eddington approximation the damping is considered to be constant and is treated as a 
free parameter during the inversion.

The macroturbulent velocity $V_{\rm mac}$ is another free parameter that accounts for other sources of broadening,
such as limited spectral resolution and velocity fields occurring at horizontal scales smaller than the resolution 
element. This is usually taken into account by convolving the synthetic Stokes profiles with a Gaussian
kernel of certain width (measured typically in km s$^{-1}$ instead of wavelength units).

In M-E atmospheres these two parameters are usually degenerated with other thermodynamic
quantities: $\eta_0$ and $\Delta \lambda_D$ (Section 2; see Orozco Su\'arez \& Del Toro Iniesta 2007). 
This means that these 4 free parameters have very similar effects on the synthetic Stokes profiles
 and thus, it is not generally possible to distinguish, for instance, between a large macroturbulence 
$V_{\rm mac}$ and a combination of a small $\eta_0$ with a large Doppler width $\Delta\lambda_D$. 
This situation is even more pronounced when we are dealing with low spectral resolution line profiles 
($\simeq$ 69 m\AA~ in HMI's case).

VFISV allows to consider all four of them as free parameters ($a$, $\eta_0$, $\Delta\lambda_D$ and $V_{\rm mac}$) but
in order to speed up the inversion process it is important to see if some of them can be skipped. Besides
the immediate benefit in the speed of the calculation of $\mathcal{H}_{\rm mod}$ (Eqs.~3-4) and its singular
values, not having $a$ as a free parameter increases also the speed at 
which we can calculate the Voigt function, since $da=0$ in Eqs.~12 and 13. In addition, having $V_{\rm mac}=0$ during the inversion 
saves precious time that is otherwise spent in expensive Fast Fourier Transforms. The question is, of course, how does this affect 
the determination of important quantities such as the three components of magnetic field vector and line-of-sight velocity of the 
plasma ?

To study this we have carried out a series of inversions over spectropolarimetric data of AR 10953 (see also Section 7).
This region was observed,  on May 3rd, 2007, with the spectropolarimeter on the Solar Optical Telescope on-board
the Japanese spacecraft Hinode (Lites et al. 2001; Kosugi et al. 2007; see also Borrero et al. 2008 and Borrero 
\& Solanki 2008 for details). Only the Fe I line at 6302.5
 \AA~ was inverted (see Sect.~3.1). We have carried out 5 different inversions. In the first one, 31 wavelength points around the selected
spectral line (with a wavelength separation of 21.5 m\AA~) were inverted having both $a$ and $V_{\rm mac}$ as free
parameters. The remaining four inversions were carried out over only 6 wavelength positions, that were obtained after
applying the HMI filters to the original profiles. In each of these three inversions the parameters $a$ and $V_{\rm mac}$
were fixed or left free.

Using the 31-wavelength inversion as a reference, we have calculated the errors in the 3 components
of the magnetic field vector: $B$ (field strength), $\gamma$ (field inclination) and $\Psi$ (field azimuth) and
LOS velocity $V_{\rm los}$ for the other 4 inversions, where only 6 wavelengths were considered. Results are displayed in
Figure 4 for $a=$free and $V_{\rm mac}=$free (solid), $a=$free and $V_{\rm mac}=0$ (dotted), $a=1$ and $V_{\rm mac}=$free
(dashed) and finally: $a=1$ and $V_{\rm mac}=0$. This figure shows that it is possible to fix $a$ and ignore the
macroturbulent velocity (making $V_{\rm mac}=0$) without significantly affecting the accuracy in the determination of the
magnetic field vector and LOS velocity. It also highlights that main source of errors is the limited spectral 
sampling: going from 31 to 6 wavelength positions.

\begin{figure}
\begin{center}
\includegraphics[width=9cm]{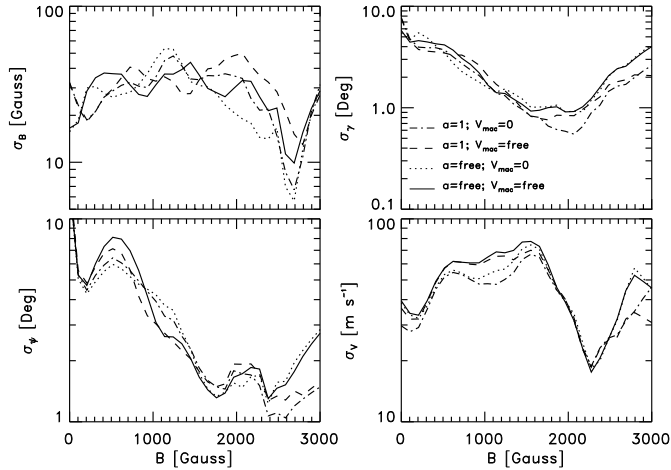}
\caption{Errors in the determination of the magnetic field strength $B$ (upper left), magnetic field inclination $\gamma$
(upper right), magnetic field azimuth $\Psi$ (lower left) and Line-of-sight velocity $V_{\rm los}$ (lower right) when
the damping $a$ and macroturbulence $V_{\rm mac}$ are inverted (solid), only $a$ is inverted (dotted), only $V_{\rm mac}$
is inverted (dashed) or neither of them are inverted (dashed-dotted).}
\end{center}
\end{figure}

\subsection{Convergence, exits and restarting}

As already mentioned in Section 2, the inversion algorithm is based on the Levenberg-Marquardt least square fitting algorithm. 
A description of this algorithm in the context of the inversion of Stokes profiles can be found
in Del Toro Iniesta (2003b) or Borrero (2004). This iterative algorithm combines the {\it steepest descent} and {\it Hessian} methods to
search for a minimum in the $\chi^2$-surface, both far away and close to this minimum. It uses a modified Hessian matrix $\mathcal{H}_{\rm mod}$, 
where the diagonal elements $\mathcal{H}_{ii}^{\rm mod}$ 
are modified with a parameter $\epsilon$ (see Eqs.~3 and 4) that weighs them more or less, as compared to the non-diagonal elements, 
whenever the minimum is believed to be far away or close, respectively. The traditional recipe for the Levenberg-Marquardt algorithm (see Press et al. 1986) 
advises to decrease or increase $\epsilon$ by one order of magnitude, depending upon a successfully ($\chi^2$ decreased) or unsuccessful ($\chi^2$ increased) 
iteration. However, in practice this recipe is not always reliable. For instance, after a number of consecutive successful iterations $\epsilon$ 
becomes very small. This implies that the modified Hessian matrix corresponds basically to the real Hessian and therefore the algorithm assumes
that we are close to the minimum. However, if after this pattern an unsuccessfully iteration appears, then $\epsilon$ starts increasing again.
The problem is that since $\epsilon$ is so small it will take a large number of extra unsuccessful iterations to bring the modified Hessian matrix to
a more diagonal topology, where the algorithm mimics the steepest descent. For this reason, in our implementation, the parameter $\epsilon$ is
not always decreased by one order of magnitude after a good iteration. An analog argument can be made whenever a successful 
iteration appears after a pattern of bad ones, and therefore we do not always increase $\epsilon$ by one order of magnitude after a bad iteration.
Table 2 details how VFISV treats the parameter $\epsilon$ depending on the success of the previous iteration.

\begin{table}
\begin{center}
\caption{Modification (new) on the $\epsilon$ parameter during the inversion with the Levenberg-Marquardt algorithm according to the previous (old)
value and whether the previous iteration was successful, $\chi^2_{\rm new} < \chi^2_{\rm old}$ , or not: $\chi^2_{\rm new} > \chi^2_{\rm old}$}
\begin{tabular}{ccc}
\hline
$\chi^2$ & $\epsilon_{\rm old}$ & $\epsilon_{\rm new}$ \\
\hline
          & $>10^{4}$ & $\epsilon_{\rm old}/100$ \\
decreases & $\ge 10^{-4}$ & $\epsilon_{\rm old}/10$ \\
          & $<10^{-4}$ & $\epsilon_{\rm old}/2$ \\
\hline
          & $<10^{-4}$ & $\epsilon_{\rm old} \times 100$ \\
increases & $>10^{4}$ & $\epsilon_{\rm old} \times 10$ \\
          & $\le 10^{4}$ & $\epsilon_{\rm old} \times 2$ \\
\hline
\end{tabular}
\end{center}
\end{table}

Unlike other inversion codes that stop iterating if $\chi^2$ is already small, VFISV performs always a fixed number of 
iterations\footnote{This number can be changed by individual users running the code. Currently set to 30.}. This is done purposedly because a small $\chi^2$ does not 
guarantee that the global minimum in the $\chi^2$-surface has been found (i.e: local minima). In order to minimize the chances of 
falling into a local minimum we prefer to restart the inversion using a randomly perturbed model, obtained out of the best model so far, and 
continue the inversion until the maximum number of iterations has been completed.

This same restarting method is also performed if: {\bf (a)} after the first 10 iterations $\chi^2$ has not decreased by at least one order 
of magnitude from its first value; {\bf (b)} $\chi^2$ has not improved after 5 consecutive iterations. The restart can happen
more than once, with increasing amplitude of the perturbations each time it happens. This level of randomization enables VFISV
to escape from local minima. Of course it can not be compared to the more elaborated randomizations performed by other methods
(simulated annealing or genetic algorithms), but at least it includes this possibility while retaining the speed of the 
Levenberg-Marquardt algorithm.

\subsection{$\chi^2$ definition and normalization}

The definition of $\chi^2$ adopted is shown in Equation 1 (Section 2). Here we discuss the meaning of $\sigma_i$ and $w_{ij}$,
which were left apart intentionally back then. $\sigma_i$ refers
to the noise level, which can be different for $I$, $Q$, $U$ and $V$ depending on the polarization calibration and the observing scheme used.
The factor $w_{ij}$ is a weighing function that is used to give different weights to the different polarization signals. $w_{ij}$ are invoked 
because the amplitude of Stokes $I$ signal is typically much larger than that of Stokes $Q$, $U$ and $V$. If $w_{ij}=1$ always, it often
happens that the derivatives of $\chi^2$, and consequently the Hessian matrix too, are dominated by the derivatives of Stokes $I$. This leads
the inversion code to focus on fitting only this Stokes parameter. For this reason most inversion codes try to compensate for this by making:
$w_{Q,U}>w_V>w_I$. For example, the SIR inversion code (Ruiz Cobo \& Del Toro Iniesta 1992) adjusts those weights such that the amplitudes
in each of observed Stokes parameter are the same. Unfortunately this procedure makes $\chi^2$ normalization pixel dependent and thus, makes it
impossible to decide whether the fit in a given pixel is better than in another one.

Other inversion codes adopt constant values for $w_{ij}$ such that $\chi^2$ values can be directly compared to each other. In 
the case of MERLIN  (Milne-Eddington gRid Linear Inversion Network) and LILIA (Lites et al. 2007) these are set by default to: $w_{Q,U} = 10 w_V =100 w_I$
(linear polarization is given 10 times more weight than circular polarization, which is given 10 times more weight than total intensity).
Since HMI will take full disk observations, it is very important that $\chi^2$ normalization is as homogeneous as possible, therefore
we have decided to follow the same approach as MERLIN, that is, employing constant weights. However, we find that the recipe of 
$w_{Q,U} = 10 w_V =100 w_I$ is somewhat unreliable when inverting pixels inside the sunspot umbra, where the amplitude of all 4 Stokes
 profiles is very similar. We have also tested the other possibility (give all 4 Stokes parameters the same weight:
$w_{Q,U} = w_V = w_I$). This approach seems to work well in active regions, but in the quiet Sun the inversion code tends to ignore
, as expected, the polarization signals and fit only the intensity $I$. This translates into a pattern for the magnetic field
strength and retrieved $\chi^2$ that closely follows the continuum intensity. These results seem to indicate that the larger
the continuum intensity the smaller weight should be given to Stokes $I$. 

After several tests (in which many maps from Hinode's spectropolarimeter have been inverted) we have come up with the following empirical recipe.
The weights in the polarization signals are left to 1: $w_Q=w_U=w_V=1$, but the weight in Stokes $I$ is calculated as:

\begin{equation}
w_I = 0.7778 \|I_{\rm cont}(\mu)-I_{\rm cont}(x,y)\|+0.3
\end{equation}

\noindent where $I_{\rm cont}(x,y)$ is the continuum intensity of a given pixel with coordinates $(x,y)$ on the solar disk. $I_{\rm cont}(\mu)$
is an estimation of the continuum intensity depending on the $\mu$ angle (see Neckel \& Labs 1994) at a reference wavelength close to 6173 \AA~:

\begin{equation}
I_{\rm cont}(\mu) = 0.33644+1.30590\mu-1.79238\mu^2+2.45050\mu^3-1.89979\mu^4+0.59943\mu^5
\end{equation}

To avoid negative values of $w_I$ whenever the continuum intensity at $(x,y)$ is very large, we are taking the absolute value of the difference
between $I_{\rm cont}(\mu)$ and $I_{\rm cont}(x,y)$ (see Equation 14). These formulas ensure that $w_I \simeq 0.3$ in quiet Sun regions, where
the continuum intensity is large, but $w_I \simeq 1$ in sunspots. Equations 14 and 15 make the $\chi^2$ normalization pixel dependent, therefore
one of our objectives to retain an homogeneous normalization is not totally fulfilled. However, considering that the only one that changes is $w_I$
(the rest are fixed to 1) and taken into account that in the worst case scenario $w_I \simeq w_{Q,U,V}$ then we can consider that $\chi^2$
normalization changes very slowly in neighboring pixels.

Finally, it is important to mention that VFISV also includes the possibility to assign different weights to different wavelength positions,
hence the $w_{ij}$ indexes. The reason for this is that the data compression and transfer is never 100 \% accurate and therefore
there will surely be some missing filtergrams (combinations of the four Stokes parameter at a given wavelength). Although we have not carried out
any experiments with it, this will be certainly be something to look into during commissioning time.

\subsection{Derivatives and Hessian}

Derivatives of $\chi^2$ with respect to the free parameters (Sect.~2.1) are the basic constituents of the divergence vector 
and Hessian matrix needed in the Levenberg-Marquart minimization algorithm (Section 2). VFISV calculates these derivatives analytically in its
synthesis module, by taking advantage of the fact that the synthetic profiles $I_{\rm syn}(\lambda)$ are analytical functions of those 
free parameters. This saves a significant amount of time as compared to other Milne-Eddington inversion codes where derivatives 
are obtained numerically by calling multiple times the synthesis model, each time slightly modifying the free parameters (e.g. MELANIE; see Lites et al. 2007).
For instance, calculating numerical derivatives for 10 free parameters require twice as many calls to the synthesis module, while 
having analytical derivatives requires only one call to the synthesis module.

Although in the case of numerical derivatives the synthesis module runs faster (since it does not calculate internally 
the derivatives), it is not fast enough to compensate the extra number of calls needed. We have confirmed this point by 
comparing the speed of the derivatives by {\bf (a)} calling the VFISV synthesis module and asking it to calculate the analytical 
derivatives internally, and {\bf (b)} calculating the numerical ones via multiple callings to the synthesis module,
while commenting the parts of the code where analytical derivatives are computed. As it can be seen in Figure 5, calculating derivatives
 analytically is almost a factor 6.4 times faster than doing it numerically.

\begin{figure}
\begin{center}
\includegraphics[width=9cm]{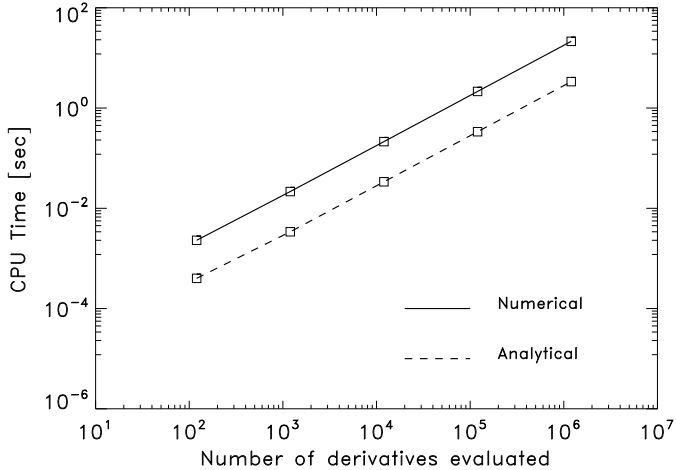}
\caption{Time need by VFISV's synthesis module to compute analytical (dashed line) and numerical derivatives (solid line). The
analytical calculation is 6.36 times faster than the numerical one.}
\end{center}
\end{figure}

Another place where we have improved significantly the speed of the code is in the way the derivatives are called. As stated in
Section 2 (see also Figure 1), some Stokes inversion codes determine the synthetic profiles $I_{\rm syn}(\mathcal{M},\lambda)$ 
and their derivatives $\partial {I_{\rm syn}(\lambda,\mathcal{M})}/ \partial{M_f}$ every time the synthesis 
module is called. It turns out that these derivatives are only needed if the iteration is successful: $\chi^2_h < \chi^2_{\rm best}$,
otherwise the derivatives employed are those corresponding to the previous best iteration. In a typical inversion about 50 \%
of the iterations are unsuccessful, meaning that about 50 \% of the derivatives (costly calculated) are unecessary. In this sense,
we have modified VFISV to call twice the synthesis module: once to determine only $I_{\rm syn}(\mathcal{M},\lambda)$, and a second
time to calculate $\partial {I_{\rm syn}(\lambda,\mathcal{M})}/ \partial{M_f}$ only if $\chi^2$ has decreased (see modified
scheme in Figure 6).

\begin{figure}
\begin{center}
\includegraphics[width=20cm,angle=-90]{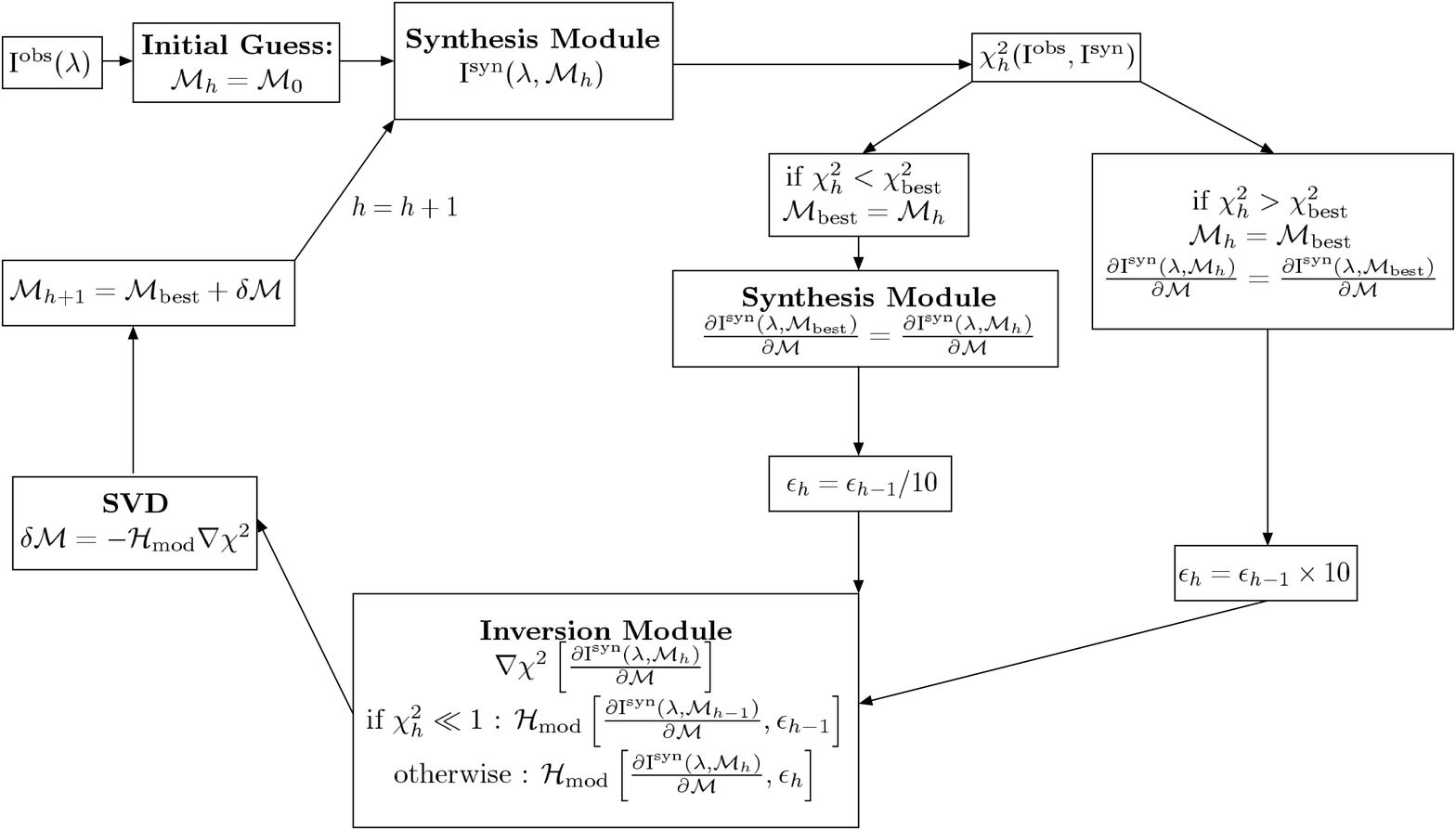}
\caption{VFISV iterative scheme for the inversion of Stokes profiles using the Levenberg-Marquadrt algorithm. The derivatives are only
computed if the $\chi^2$ at iteration $h$ is smaller than any previous value. Otherwise the derivatives from the previous best iteration
are used. In addition the Hessian matrix is not calculated by the inversion module if $\chi^2_h \ll 1$.}
\end{center}
\end{figure}

Another computationally expensive part of VFISV is the calculation, at each iteration step, of the modified Hessian matrix.
To avoid this, VFISV does not calculate the modified Hessian matrix whenever we are close to a minimum. Instead, VFISV reuses
the Hessian matrix from the previous iteration (if that one was also close to the minimum). This can be done because
close to the minimum the modified Hessian matrix represents the curvature of the $\chi^2$-surface, and this curvature
is approximately the same around the minimum. This is also indicated in Figure 6 (cf. Figure 1).

\section{Initialization module}%

A very critical part in the inversion of Stokes profiles is the selection of a suitable set of model parameters to start the
iterative process: $\mathcal{M}_0$ (see Figure 1). Although in general these inversion codes are robust enough (Westendorp Plaza et al. 1998) to guarantee
that similar results are achieved regardless of the starting point (specially if the inversion includes some randomization; see Section 4.2),
the closer the initial model is to the solution, the fewer iterations will be needed. In a problem such as ours, where speed is critical, 
it is mandatory to reduce these iterations by determining a good initial model. Of course, whatever method is used to initialize the
full inversion, it must require only a very small fraction of the total inversion time.

VFISV uses two different methods to calculate an initial guess model. The first method, based on Artificial Neural Networks, is applied 
to determine $B$ and $\gamma$ whenever the total polarization signal is above 6 \%. The second technique, based on a combination of the 
magnetograph formula and the Weak Field Approximation (Jefferies \& Mickey 1991), is also applied to $B$ and $\gamma$ if the polarization level 
is below 6 \%. This threshold was selected by determining how close each of two different initializations was to the final solution
obtained through the full inversion (see Figure 7).

\begin{figure}
\begin{center}
\includegraphics[width=7cm]{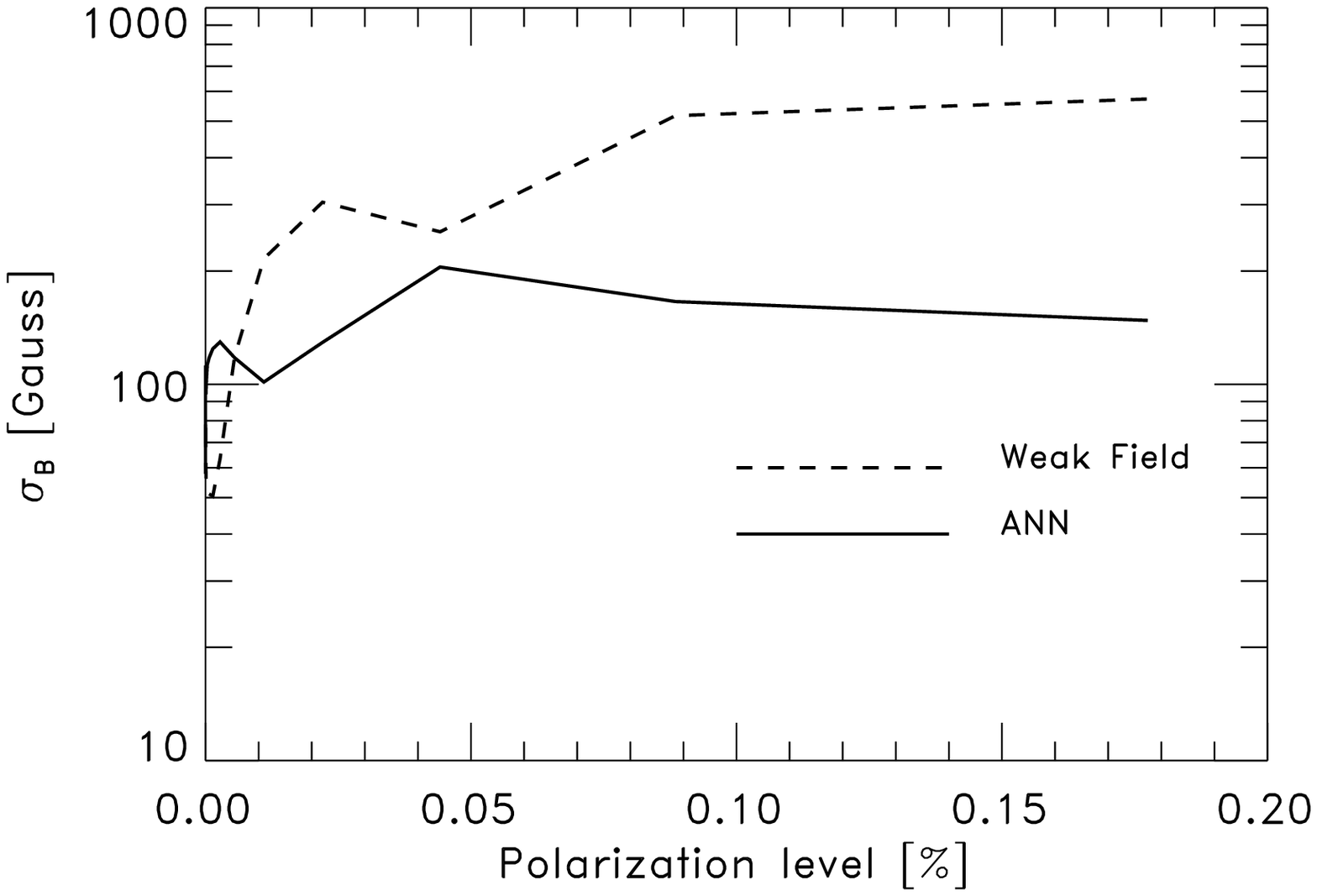} \\
\includegraphics[width=7cm]{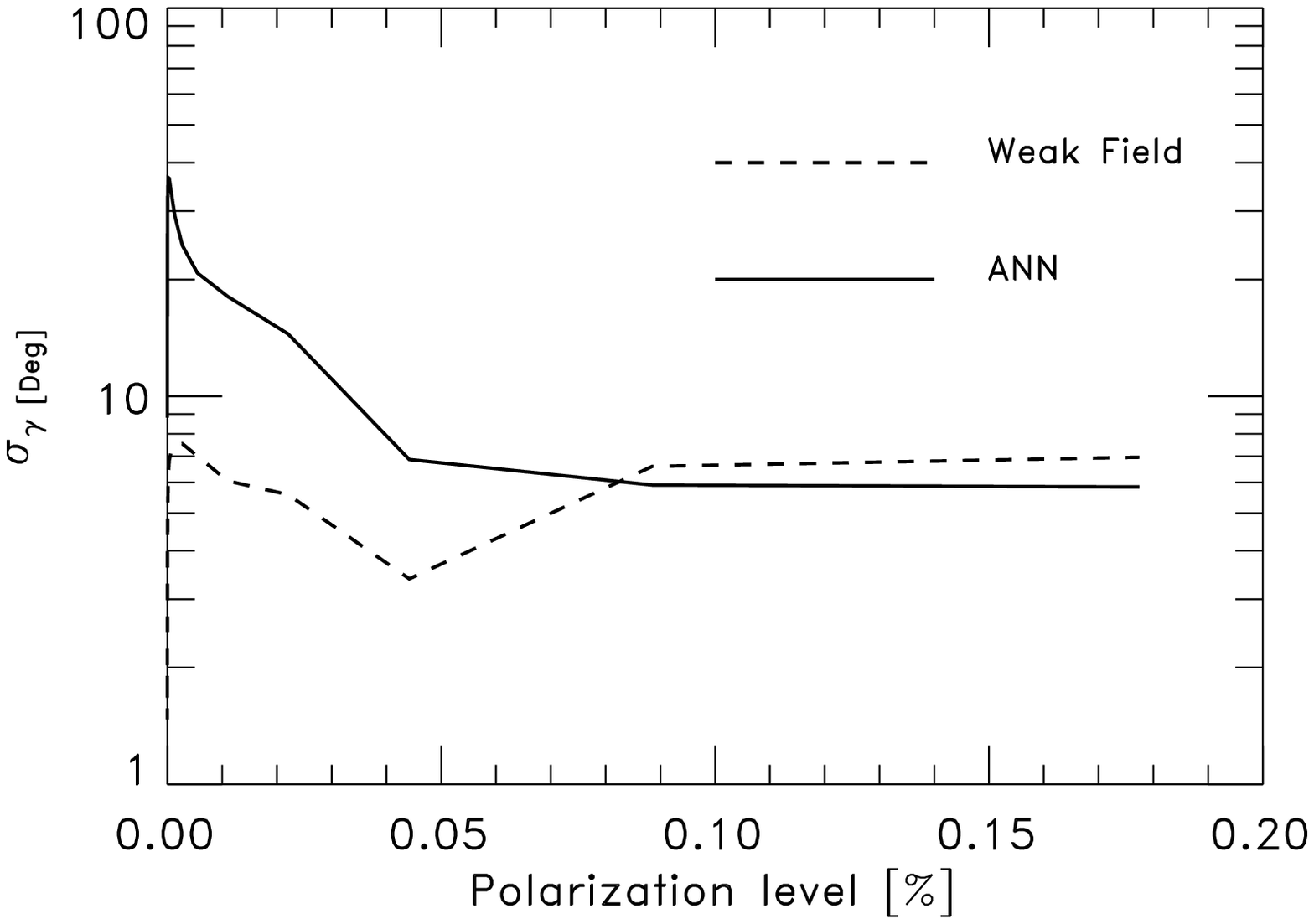}
\caption{{\it Top panel}: Standard deviation in the magnetic field strength ($\sigma_B$) between the initialized value and the
final one obtained from the full inversion, as a function of the polarization level in the observed Stokes profiles. The initializations
were obtained with Artificial Neural Networks (solid line) and the Weak Field Approximation (dashed line). {\it Bottom panel}: same but
for the inclination of the magnetic field $\gamma$.}
\end{center}
\end{figure}

The azimuthal angle $\Psi$ is always determined using the approach proposed by Auer et al. (1977).
We find this method more reliable than applying the well known formula: $\Psi = \frac{1}{2} \tan^{-1} {\frac{U}{Q}}$. It is also far
superior than results from Neural Networks. Finally, the line-of-sight velocity $V_{\rm los}$ is always determined using Artificial Neural Networks.

Artificial Neural Networks for $B$, $\gamma$ and $V_{\rm los}$ were individually trained using a back-propagation method (Bishop 1994). We
employed a 3-layered net with 30 neurons per layer. Non-linear transformations were performed between layers. For each of the training model
 parameters two different sets of profiles were created:  the training and the control set (with 250000 Stokes profiles each). The Stokes profiles 
were obtained by producing synthetic profiles using the synthesis module of the SIR inversion code and randomizing all its properties: $V_{\rm mac}$, 
$T(\tau)$ (temperature stratification), $B$, $\gamma$, $\Phi$ and $V_{\rm los}$. The profiles were convolved with the theoretical HMI filter functions 
to obtain 6 wavelength positions. Finally, we also added random noise to the level of $10^{-3}$ to the synthetic Stokes profiles.

Two different sets are needed because the error in the training set always drops to zero with time (the ANN specializes)
, but in the control set it only decreases initially. The optimum moment to stop the learning process
is when the error in the control set starts to increase, as this indicates that Neural Net is loosing generality and is
becoming too specialized. Figure 8 (top panel) shows and example of how the error in the determination
of the line-of-sight velocity $V_{\rm los}$ decreases with the training time (in the control set). After 24 hours
of training $\sigma_v$ started to increased again (see vertical dashed line). At that point the error was $\sigma_v \simeq 93$ m s$^{-1}$. Figure 8 (bottom panel)
shows a scatter plot of the velocity in the control set of Stokes profiles and the velocity determined by the Neural Network.

\begin{figure}
\begin{center}
\includegraphics[width=7cm]{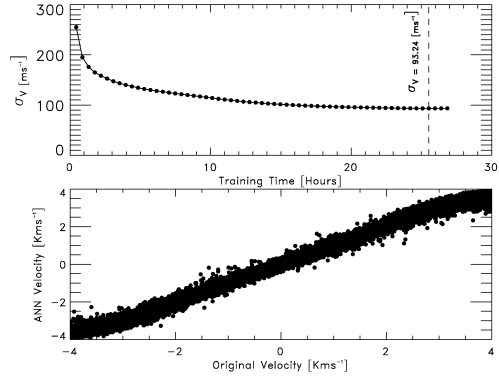}
\caption{{\it Top panel}: error in the determination of the line-of-sight velocity $\sigma_v$, using neural networks, as a function of the training time.
{\it Bottom panel}: scatter plot of the real $V_{\rm los}$ versus the one determined using ANNs.}
\end{center}
\end{figure}

\section{VFISV profiling and speed}%

In this section we present Tables 3-5, where we show the time spent in each of the VFISV modules and subroutines, as well as a brief description. These
results correspond to the different levels of optimization described in the previous sections. They were obtained after inverting a sample of 25 FITS 
files, using 1 single cpu, from Hinode/SP data (a total of 25600 Stokes profiles).

\begin{table}
\begin{center}
\caption{Profiling of the Very Fast Inversion of the Stokes Vector (VFISV) without any optimization. The inversion run at a speed of {\bf 479.75 profiles/second/cpu}.
The indexes for each subroutine correspond to $1$: Computes synthetic Stokes vector and its derivatives; $2$: Computes singular values of modified Hessian matrix
; $3$: Computes elements of absorption matrix and its derivatives; $4$: Computes the modified Hessian matrix $\mathcal{H}$; $5$: Computes Voigt and Voigt-Faraday 
functions; $6$: Main driver of the program; $7$: Computes divergence of $\chi^2 $vector: $\nabla\chi^2$; $8$: Normalizes derivatives of Synthetic Stokes profiles;
$9$: Solves linear system of Equations (see Eq.~5); $10$: Determines initial guess model $\mathcal{M}_0$ (see Figures 1 and 6); $11$: Determines $\chi^2$ at 
each iteration step; $12$: Checks for overflows, NaN and Inf.}
\begin{tabular}{|ccccc|}
\hline
Subroutine Name & Module & Time [\%] & Time [sec] & Calls \\
\hline\hline
Forward$^1$ & Synthesis & 38.82 & 18.14 & 742400 \\
Svdcmp$^2$ & Inversion & 21.28 & 9.94 & 742400 \\
Gethess$^4$ & Inversion & 12.07 & 5.64 & 742400 \\
Absmat$^3$ & Synthesis & 10.10 & 4.72 & 742400 \\
Voigt$^5$ & Synthesis & 8.53 & 3.99 & 2227200 \\
VFISV$^6$ & MAIN & 2.44 & 1.14 & 1\\
Getdiv$^7$ & Inversion & 1.78 & 0.83 & 742400 \\
Normalize$_{-}$dsyn$^8$ & Inversion & 1.26 & 0.59 & 530336  \\
Svbksb$^9$ & Inversion & 0.83 & 0.39 & 742400  \\
Getchi2$^{11}$ & Inversion & 0.30 & 0.14 & 742400  \\
Dnanchk$^{12}$ & MAIN & 0.22 & 0.11 & 19302400 \\
Guess$^{10}$ & MAIN & 0.21 & 0.10 & 25600 \\
\hline
\end{tabular}
\end{center}
\end{table}

Between Table 3 and Table 4 the only change was the optimization in the calculation of the Voigt and Voigt-Faraday functions, which was
originally computed using the algorithm by Hui et al. (1977) and changed into a second order Taylor expansion. In the first case, the subroutine
{\it Voigt} took 3.99 seconds, whereas in the second case {\it Voigt$_{-}$taylor} needed only 1.30 seconds. This is very close to the factor
3.4 already foreseen in Section 3.2. Note also that, as requested in Section 5, the initialization performed in subroutine {\it Guess}
needs a negligible amount of time compared to the rest of the inversion.

In Table 5 we show a similar profiling where we now also include the optimization in the derivatives and Hessian matrix described in
section 4.4. Note that the Synthesis module (subroutines {\it Forward} and {\it Absmat}) are now called about 50 \% more times (see Fig.~6).
However, the total amount of time they run was significantly less (about 50 \% faster). In addition, the Hessian matrix was calculated much
less often, which translated into a smaller time running subroutine {\it Gethess}.

\begin{table}
\begin{center}
\caption{Profiling of the Very Fast Inversion of the Stokes Vector (VFISV) including optimization for the Voigt function (see Sect.~3.2). 
The inversion run at a speed of {\bf 514.66 profiles/second/cpu}.}
\begin{tabular}{|ccccc|}
\hline
Subroutine Name & Module & Time [\%] & Time [sec] & Calls \\
\hline\hline
Forward$^1$ & Synthesis & 42.86 & 18.71 & 742400 \\
Svdcmp$^2$ & Inversion & 21.22 & 9.26 & 742400 \\
Absmat$^3$ & Synthesis & 12.74 & 5.56 & 742400 \\
Gethess$^4$ & Inversion & 10.67 & 4.67 & 742400 \\
Voigt$_{-}$Taylor$^5$ & Synthesis & 2.98 & 1.30 & 2227200 \\
VFISV$^6$ & MAIN & 2.43 & 1.06 & 1\\
Getdiv$^7$ & Inversion & 1.82 & 0.80 & 742400 \\
Normalize$_{-}$dsyn$^8$ & Inversion & 1.26 & 0.55 & 514146  \\
Svbksb$^9$ & Inversion & 1.21 & 0.53 & 742400  \\
Getchi2$^{11}$ & Inversion & 0.39 & 0.17 & 742400  \\
Guess$^{10}$ & MAIN & 0.21 & 0.09 & 25600 \\
Dnanchk$^{12}$ & MAIN & 0.07 & 0.03 & 19302400 \\
\hline
\end{tabular}
\end{center}
\end{table}

\begin{table}
\begin{center}
\caption{Profiling of the Very Fast Inversion of the Stokes Vector (VFISV) including optimization for the Voigt function (see Sect.~3.2)
and optimization in the derivatives and Hessian matrix (see Sect.~4.4). The inversion run at a speed of {\bf 686.24 profiles/second/cpu}.}
\begin{tabular}{|ccccc|}
\hline
Subroutine Name & Module & Time [\%] & Time [sec] & Calls \\
\hline\hline
Forward$^1$ & Synthesis & 33.80 & 10.86 & 1055429 \\
Svdcmp$^2$ & Inversion & 28.99 & 9.31 & 742400 \\
Absmat$^3$ & Synthesis & 14.85 & 4.77 & 1055429 \\
Gethess$^4$ & Inversion & 5.42 & 1.74 & 296833 \\
Voigt$_{-}$Taylor$^5$ & Synthesis & 4.90 & 1.58 & 3166287 \\
VFISV$^6$ & MAIN & 3.42 & 1.10 & 1\\
Getdiv$^7$ & Inversion & 2.62 & 0.84 & 742400 \\
Svbksb$^9$ & Inversion & 1.15 & 0.37 & 742400  \\
Normalize$_{-}$dsyn$^8$ & Inversion & 0.84 & 0.27 & 313029  \\
Getchi2$^{11}$ & Inversion & 0.56 & 0.18 & 742400  \\
Guess$^{10}$ & MAIN & 0.39 & 0.13 & 25600 \\
Dnanchk$^{12}$ & MAIN & 0.06 & 0.02 & 19302400 \\
\hline
\end{tabular}
\end{center}
\end{table}

\section{Tests with Hinode/SP data}%

In this section we present results from the application of VFISV to spectropolarimetric data from Hinode/SP.
We have used here two different data sets. They correspond to observations carried out on November 14, 2006
and May 3, 2007 respectively. Both datasets encompass active regions AR 10923 and 10953, which were located
at around $\mu=0.93$. The second dataset has already been used in our experiments in Sections 4.1 and 6. They
are comprised by a set of 538 and 1000 fits files, therefore having 550,912 and 1,024,000 Stokes profiles, respectively.
Since Hinode/SP observes the Fe I lines at 6301.5 and 6302.5 \AA~ we have trimmed the observations to use only the
second spectral line (see Section 3.1). After that, we convolved the original observations with the theoretical
filter profiles from HMI, to reduce the number of wavelength positions to 6\footnote{The VFISV version 
(currently 1.03) that has been made public to the community has the capability of inverting polarimetric data
with any spectral resolution, but here we restrict ourselves to using 6 wavelength positions to simulate HMI's case.}.

Each map has been inverted using a Quad-core (4-CPUs) Intel Xeon machine at 2.66 GHz. The first map was inverted in 
227.13 seconds, whereas the second one was ready after 405.88 seconds. This translates into a speed of 606.38 and
630.73 profiles/sec/cpu. Thus, inverting 22500 profiles/seconds (as requested by HMI; see Section 1) is possible
with only 35-40 CPUs.

Maps of the magnetic field strength for these two regions are presented in Figure 9. For better visualization only the regions
around the two sunspots are displayed (the original maps are 86"$\times$164" and 163"$\times$164"). We have also tested VFISV
with more than 1000 maps available from the Hinode/SP database that have been verified for Level-1 (Bruce Lites, {\it private
communication}). We have found that VFISV yields adequate results for all kind of structures: network, plage, sunspots, etc.
The main issues that we have found are:

\begin{itemize}
\item We have observed systematic errors in the umbra of those sunspots, when these are located very close to the limb ($\mu < 0.15$).
\item When inverting using a filling factor ($\alpha_{\rm mag}$ free) about 1-2 \% of pixels in the very quiet Sun regions
(e.g. internetwork) yield inadequate results. These are characterized by very large magnetic field strength, $B > 4000$ Gauss, 
and very small magnetic filling factors: $\alpha_{\rm mag}<0.05$.
\item In sunspots where the magnetic field is extremely high in the umbra, $B > 4000$ Gauss, we observe saturation effects in the magnetic field 
strength whenever we invert only 6 wavelength positions. This effect disappears when using the full spectral profile.
\end{itemize}

\begin{figure}
\begin{center}
\includegraphics[width=14cm]{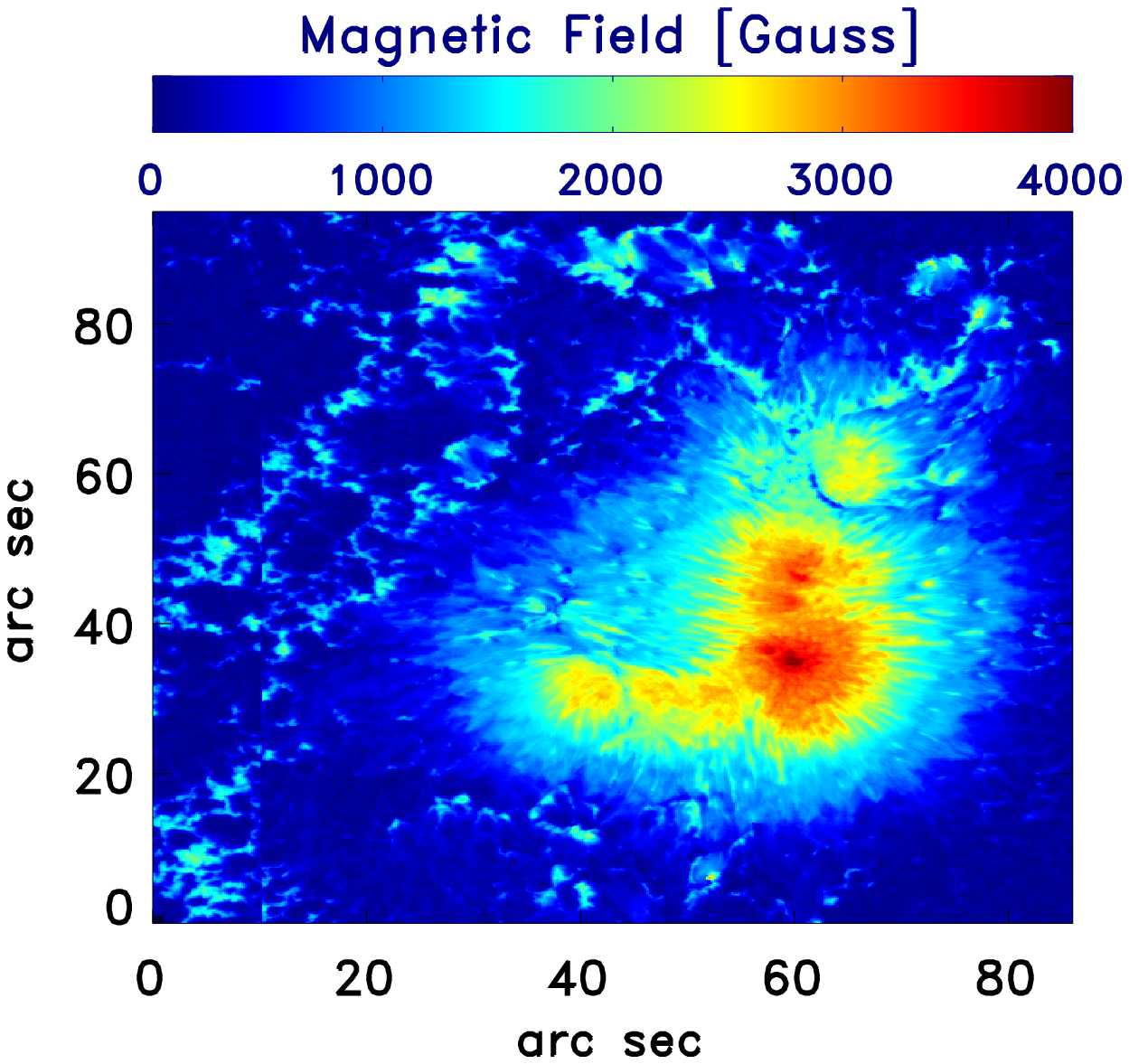} \\
\includegraphics[width=14cm]{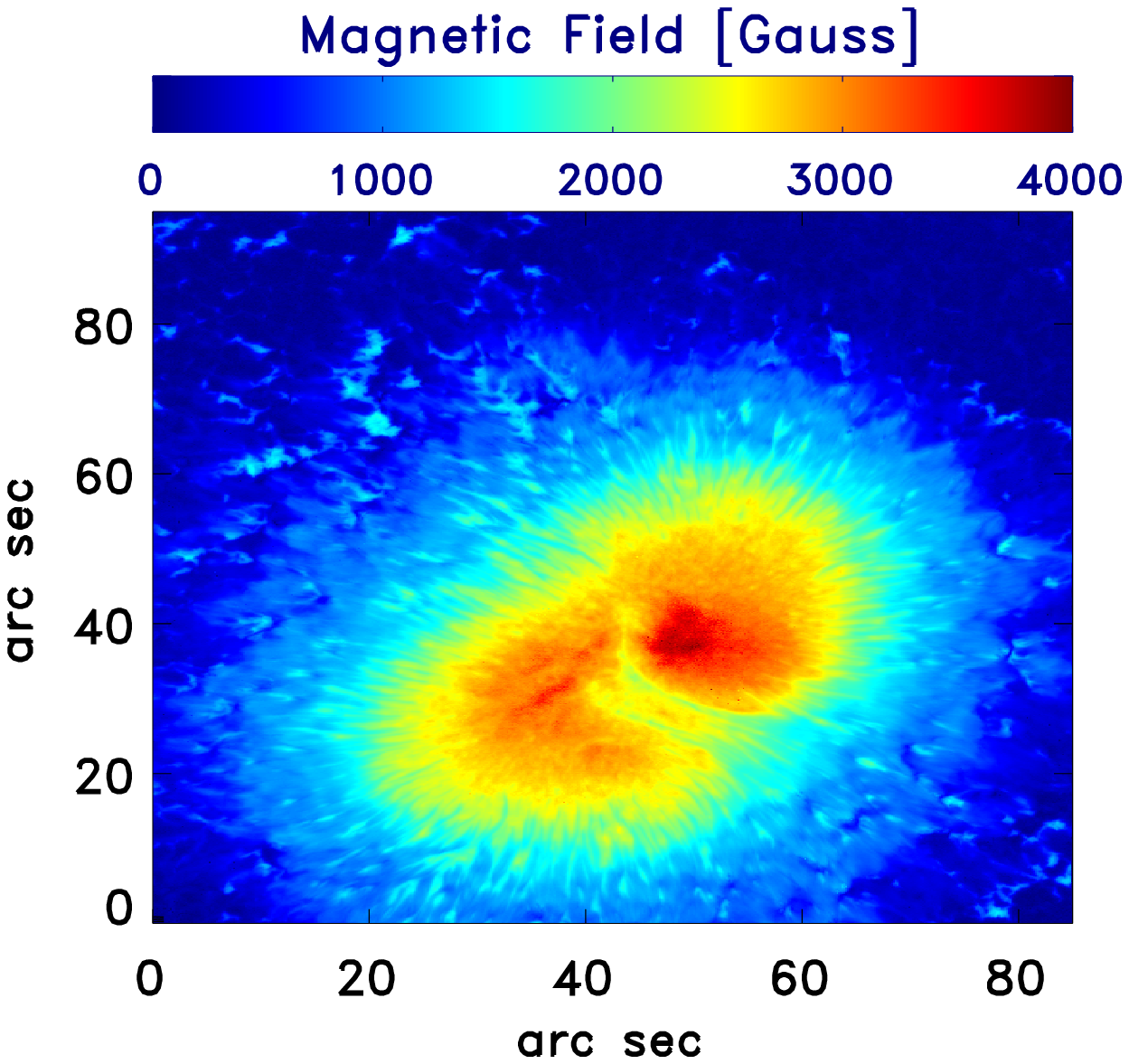}
\caption{Maps of the magnetic field strength obtained from the inversion of Hinode/SP data using the VFISV inversion code. Top panel
shows AR 10953 and bottom panel shows AR 10923.}
\end{center}
\end{figure}

\section{Conclusions}%

We have developed an inversion code for the polarized radiative transfer equation. The code's name is VFISV: Very Fast
Inversion of the Stokes Vector. This code assumes that the properties of the Solar Photosphere are well described by the Milne-Eddington 
approximation. It will be employed to routinely invert polarimetric data from the Heliseismic and Magnetic Imager (HMI) that will fly 
on-board of the Solar Dynamics Observatory (SDO) in June 2009/2010. The code is written in Fortran 90 and parallelized with MPICH-2. 
We have introduced a number of improvements in this code that makes it able to achieve an inversion speed of about 600 pixels per seconds per 
CPU. At this pace it will be possible to provide Solar full-disk maps (with 1" resolution) of the Solar magnetic field vector every 10 minutes, 
using less than 40 CPUs.

The code has been tested with observations from the spectropolarimeter onboard Hinode. It is freely available for download through the 
Community Spectro-polarimetric Analysis Center initiative at the High Altitude Observatory and National Center for Atmospheric
Research (Lites et al. 2007; http://www.hao.ucar.edu/projects/csac/).


\begin{thebibliography}{}
\bibitem{auer}
Auer, L.H., House, L.L. \& Heasley, J.N. 1977: {\it Solar Phys}, 55, 47-61
\bibitem{baur1}
Baur, T.G., Elmore, D.E., Lee, R.H, Qerfeld, C.W. \& Rogers, S.R. 1981: {\it Solar Phys}, 70, 395
\bibitem{luis1}
Bellot Rubio, L.R. 2006:  ASP Conference Series, 358, 107. Proceedings of the Solar Polarization 
Workshop 4 in Boulder, Colorado, USA. Eds: Roberto Casini and B.W.~Lites
\bibitem{bishop}
Bishop, C.M. 1994: Neural Networks for Pattern Recognition. Oxford University Press, 1994.
\bibitem{borrero1}
Borrero, J.M., Tomczyk, S., Norton, A., Darnell, T., Schou, J., Scherrer, P., Bush, R.
\& Liu, Y. 2007: {\it Solar Phys}, 240, 177
\bibitem{borrero2}
Borrero, J.M., Lites, B.W. \& Solanki, S.K. 2008: A\&A, 481, L13-L16
\bibitem{borrero3}
Borrero, J.M. \& Solanki, S.K. 2008: ApJ, 687, 668-677
\bibitem{borrero4}
Borrero, J.M.: 2004, PhD Thesis, G\"ottingen Univertity (Germany), Published by Copernicus GmbH.
ISBN 3-936596-33-0
\bibitem{jc1}
Del Toro Iniesta, J.C. 2003a: Astronomische Nachrichten, 324, no.4, 383
\bibitem{jc2}
Del Toro Iniesta, J.C. in Introduction to Spectropolarimetry. Cambridge University Press, 2003b. ISBN: 0521818273
\bibitem{jefferies}
Jefferies, J.T. \& Mickey, D.L. 1991: ApJ, 372, 694-702
\bibitem{hui}
Hui, A.K., Armstrong, B.H. \& Wray, A.A. 1977: JQSTR, 18, 509-516
\bibitem{hinode}
Kosugi, T., Matsuzaki, K., Sakuo, T. et al. 2007: {\it Solar Physics}, 243, 3
\bibitem{kendra}
Letchworth, K.L. \& Benner, D.C 2007: JQSRT, 107, 173-192
\bibitem{bruce}
Lites, B.W., Elmore , D.F. \& Streander, K.V. 2001: ASP Conf. Series, vol 236. Advanced
Solar Polarimetry. Ed. M.~Sigwarth (San Francisco: ASP)
\bibitem{nave}
Nave, G., Johansson, S., Learner, R.C.M. \& Thorne, P. 1994: ApJS, 94, 221-459
\bibitem{neckel}
Neckel, H. \& Labs, D. 1994: {\it Solar Phys}, 153, 91-114
\bibitem{landolfi}
Landolfi, M. \& Landi Degl'Innocenti, E. 1982: {\it Solar Phys}, 78, 355
\bibitem{bruce1}
Lites, B., Casini, R., Garcia, J. \& Socas-Navarro, H. 2007: Memorie della Societa Astronomica
Italiana, 78, 148
\bibitem{david}
Orozco Su\'arez \& Del Toro Iniesta, J.C.: 2007, A\&A, 462, 1137-1145
\bibitem{press}
Press, W., Flannery, B., Teukolsky, S. \& Vetterling, W. 1986: Numerical Recipes - The Art
of Scientific Computing. Cambridge: Cambridge University Press.
\bibitem{rees00}
Rees, D.E., L\'opez Ariste A., Thatcher, J. \& Semel, M. 2000: A\&A, 355, 759.
\bibitem{basilio1}
Ruiz Cobo, B. 2006: Modern solar facilities - advanced solar science, Proceedings of a Workshop held at G\"ottingen 
September 27-29, 2006 ISBN 978-3-938616-84-0. Published by Universit\"atsverlag Göttingen. 
\bibitem{andy1}
Skumanich, A. \& Lites, B.W. 1987: ApJ, 322, 473
\bibitem{hector1}
Socas-Navarro, H., L\'opez Ariste, A. \& Lites, B.W. 2001: ApJ, 553, 949
\bibitem{hector2}
Socas-Navarro, H.: 2003, Neural Networks, 16, 355
\bibitem{hector3}
Socas-Navarro, H.: 2005, ApJ, 621, 545
\bibitem{unsold}
Unsold, A.: 1955, Physik der Sternatmospharen. Berlin, Springer.
\bibitem{carlos1}
Westendorp Plaza, C., del Toro Iniesta, J.C., Ruiz Cobo, B., Mart{\'\i}nez Pillet, V., Lites, B.W. \& Skumanich, A. 1998: ApJ, 494, 453
\bibitem{axel}
Wittmann, A.: 1974, {\it Solar Phys}, 35, 11-29
\end{thebibliography}
\end{document}